\documentstyle[emulateapj,psfig,apjfonts]{article}

\def\axp{AX~J1845--0258}
\def\ax2{AX~J184453.3--025642}
\def\snr{G29.6+0.1}
\def\etal{{\rm et~al.\ }}
\def\HI{H\,{\sc i}}

\def\kms{km~s$^{-1}$}

\lefthead{GAENSLER, GOTTHELF AND VASISHT}
\righthead{A SNR COINCIDENT WITH THE SLOW X-RAY PULSAR \axp}

\begin{document}
\title{A supernova remnant coincident with the slow X-ray pulsar \axp}
\submitted{(To appear in {\em The Astrophysical Journal Letters})}
\author{B. M. Gaensler\altaffilmark{1,2}, E. V. Gotthelf\altaffilmark{3}
and G. Vasisht\altaffilmark{4}}

\altaffiltext{1}
{Center for Space Research, Massachusetts Institute of Technology,
70 Vassar Street, Cambridge, MA 02139; bmg@space.mit.edu}
\altaffiltext{2}{Hubble Fellow}
\altaffiltext{3}{Columbia Astrophysics Laboratory, Columbia University, 550 West 120th Street,
New York, NY 10027; evg@astro.columbia.edu}
\altaffiltext{4}{Jet Propulsion Laboratory, California Institute of Technology, 4800 Oak Grove
Drive, Pasadena, CA 91109; gv@astro.caltech.edu}

\begin{abstract}

We report on Very Large Array observations in the direction of the
recently-discovered slow X-ray pulsar \axp. In the resulting images, we
find a $5'$ shell of radio emission; the shell is linearly polarized
with a non-thermal spectral index. We class this source as a previously
unidentified, young ($<8000$~yr), supernova remnant (SNR), \snr, which
we propose is physically associated with \axp. The young age of
\snr\ is then consistent with the interpretation that anomalous X-ray
pulsars (AXPs) are isolated, highly magnetized neutron stars
(``magnetars'').  Three of the six known AXPs can now be associated
with SNRs; we conclude that AXPs are young ($\la$10\,000~yr) objects,
and that they are produced in at least 5\% of core-collapse supernovae.

\end{abstract}


\section{Introduction}

It is becoming increasingly apparent that isolated neutron stars come in
many flavors besides traditional radio pulsars.  In recent years, the
neutron star zoo has widened to include $\sim$10 radio-quiet neutron
stars (\cite{bj98}), six anomalous X-ray pulsars (AXPs; \cite{mer99})
and four soft $\gamma$-ray repeaters (SGRs; \cite{kou99}).  There is
much uncertainty and debate as to the nature of these sources; one way
towards characterizing their properties is through associations with supernova
remnants (SNRs).  An association with a SNR gives an independent estimate
of a neutron star's age and distance, while the position of the neutron
star with respect to the SNR's center can be used to estimate the
transverse space velocity of the compact object.

A case in point are the AXPs. Some authors propose that the AXPs
are accreting systems (\cite{vtv95}; \cite{ms95};  
\cite{gaw97}), while others argue that AXPs are ``magnetars'', isolated
neutron stars with very strong magnetic fields, $B \ga 10^{14}$~G
(\cite{td96b}; \cite{hh97}; \cite{mel99}).  However, the association
of the AXP~1E~1841--045 with the very young ($\la$2~kyr) SNR~G27.4+0.0
(\cite{vg97}) makes the case that 1E~1841--045 is a young object. Assuming
that the pulsar was born spinning quickly, it is difficult to see how
accretion could have slowed it down to its current period in such a short
time. This result thus favors the magnetar model for 1E~1841--045, and indeed
the magnetic field inferred from its period and period derivative,
and assuming standard pulsar spin-down, is
$B \approx 8 \times 10^{14}$~G.

\axp\ (also called AX~J1844.8--0258) 
is a 6.97~sec X-ray pulsar, found serendipitously in
an {\em ASCA}\ observation of the (presumably unassociated)
SNR~G29.7--0.3 (\cite{gv98}, hereafter GV98; \cite{tkk+98}, hereafter
T98). The long pulse period, low Galactic latitude and soft spectrum of
\axp\ led GV98 and T98 to independently propose that this source is an
AXP (a conclusion which still needs to be confirmed through measurement
of a period derivative). The high hydrogen column density
inferred from photoelectric absorption ($N_H \approx 10^{23}$~cm$^{-2}$)
suggests that \axp\ is distant; T98 put it in the Scutum arm, with
a consequent distance of 8.5~kpc, while GV98 nominate 15~kpc.

Because \axp\ was discovered at the very edge of the {\em ASCA}\ GIS
field-of-view, its position from these observations could only be
crudely estimated, with an uncertainty of $\sim3'$.  
A subsequent
(1999~March) 50~ks on-axis {\em ASCA}\ observation has since been carried out
(\cite{vgtg99}).  No pulsations are seen in
these data, but a faint point source, \ax2, is detected within
the error circle for \axp.
Vasisht et al.\ (1999) determine an accurate position for \ax2,
and argue that it corresponds to \axp\ in a quiescent state.
Significant variations in the flux density of \axp\ were also reported
by T98.

The region containing \axp\ has been surveyed at 1.4~GHz as part of the
NVSS (\cite{ccg+98}). An image from this survey shows a $\sim5'$ shell
near the position of the pulsar.  We here report on multi-frequency
polarimetric observations of this radio shell, at substantially
higher sensitivity and spatial resolution than offered by the NVSS.
Our observations and analysis are described in \S\ref{sec_obs},
and the resulting images are presented in \S\ref{sec_results}.  In
\S\ref{sec_discuss} we argue that the radio shell coincident with \axp\
is a new SNR, and consider the likelihood of an association between the
two sources.

\section{Observations and Data Reduction}
\label{sec_obs}

Radio observations of the field of \axp\ were made with the
D-configuration of the Very Large Array (VLA) on 1999 March 26. The total
observing time was 6~hr, of which 4.5~hr was spent observing in the 5~GHz
band, and the remainder in the 8~GHz band. 5~GHz observations consisted
of a 100~MHz bandwidth centered on 4.860~GHz; 8~GHz observations were
similar, but centered on 8.460~GHz.  Amplitudes were calibrated by
observations of 3C~286, assuming flux densities of 7.5 and 5.2~Jy at
5~GHz and 8~GHz respectively. Antenna gains and instrumental polarization
were calibrated using regular observations of MRC~1801+010.  Four Stokes
parameters (RR, LL, RL, LR) were recorded in all observations.  To cover
the entire region of interest, observations were carried out in a mosaic
of 2 (3) pointings at 5 (8)~GHz.

Data were edited and calibrated in the {\tt MIRIAD} package.  In total
intensity (Stokes~$I$), mosaic images of the field were formed using
uniform weighting and maximum entropy
deconvolution.  The resulting images were then corrected for both the
mean primary beam response of the VLA antennas and the mosaic pattern. The
resolution and noise in the final images are given in Table~\ref{tab_obs}.
Images of the region were also formed in Stokes~$Q$, $U$ and $V$.  These
images were made using natural weighting to give maximum sensitivity, and
then deconvolved using a joint maximum entropy technique (\cite{sbd99}).
At each of 5 and 8~GHz, a linear polarization image $L$ was formed from
$Q$ and $U$.  Each $L$ image was clipped where the polarized emission
or the total intensity was less than 5$\sigma$.

In order to determine a spectral index from these data, it is important to
ensure that the images contain the same spatial scales.  We thus
spatially filtered each total intensity image (see \cite{gbm+98}),
removing structure on scales larger than $5\arcmin$ and smoothing each
image to a resolution of $15\arcsec$. The spatial distribution of
spectral index was then determined using the method of ``T--T''
(temperature-temperature) plots (\cite{tpkp62}; \cite{gbm+98}).
%
%

\section{Results}
\label{sec_results}

\begin{figure*}[htb]
\centerline{\psfig{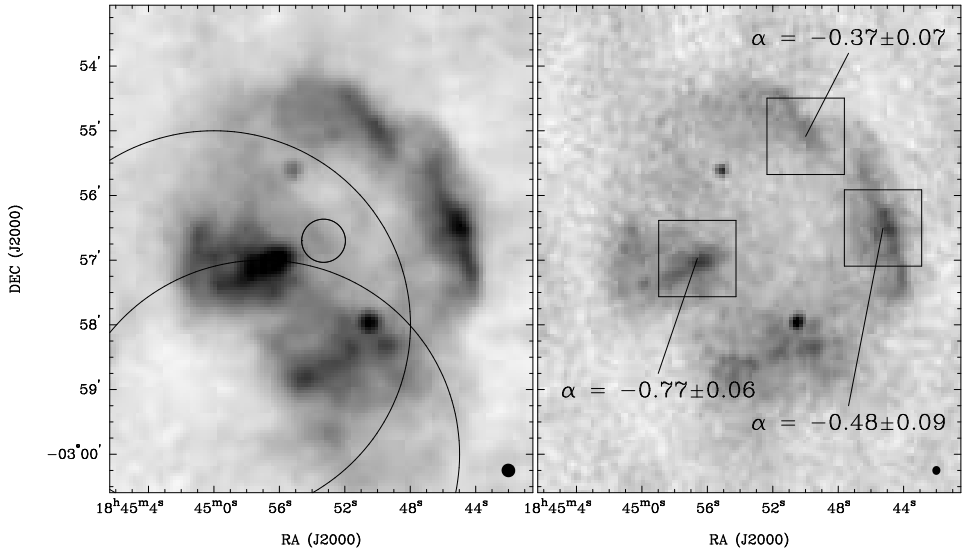}}
\caption{VLA images of \snr. On the left is a 5~GHz image, corrected
for primary beam attenuation and shown with a grey scale range of
--0.4 to 2.5~mJy~beam$^{-1}$. The large circles correspond to
the position estimates for \axp\ of GV98 (upper) and T98 (lower)
respectively; the small circle marks
the position of \ax2\ (\cite{vgtg99}).
On the right is a 8~GHz image,
uncorrected for primary beam effects; the grey
scale range is --0.2 to 1.3~mJy~beam$^{-1}$. The boxes correspond
to regions in which a spectral index, $\alpha$ ($S_\nu \propto
\nu^{\alpha}$), could
be calculated. The synthesized beams are shown at lower right.}
\label{fig_snr}
\end{figure*}

Total intensity images of the region are shown in Figure~\ref{fig_snr}.
At both 5 and 8~GHz, a distinct shell of emission is seen, which
we designate \snr; observed properties are given in Table~\ref{tab_obs}. 
The shell is clumpy, with a particularly bright
clump on its eastern edge. In the east the shell is quite thick (up
to 50\% of the radius), while the north-western rim is brighter and narrower.
Two point sources can be seen within the shell interior. At 5~GHz,
the shell appears to be sitting upon a plateau of diffuse extended emission;
this emission is resolved out at 8~GHz.

Significant linear polarization at 5~GHz is seen from
much of the shell, particularly in the two brightest parts of
the shell on the eastern and western edges. Where
detected, the fractional polarization is 2--20\%. At 8~GHz,
linear polarization is seen only from these two regions, with fractional
polarization 5--40\%. No emission was detected in
Stokes $V$, except for instrumental effects associated with the offset
of the VLA primary beam between left- and right-circular polarization.

Meaningful T--T plots were obtained for three regions of the SNR,
as marked in Figure~\ref{fig_snr}; the spectral index, $\alpha$ ($S_\nu 
\propto \nu^{\alpha}$), for each region is marked. There appear
to be distinct variations in spectral index around the shell,
but all three determinations fall in the range $-0.7 \la \alpha \la -0.4$.

Two point sources are visible within the field. The more northerly of
the two is at $18^{\rm h}44^{\rm m}55\farcs11$,
$-02\arcdeg55\arcmin36\farcs9$ (J2000), with $S_{\rm 5\,GHz} =
0.8\pm0.1$~mJy and $\alpha = +0.5 \pm 0.3$, while the other is at
$18^{\rm h}44^{\rm m}50\farcs59$, $-02\arcdeg57\arcmin58\farcs5$ (J2000)
with $S_{\rm 5\,GHz} = 2.0\pm0.3$~mJy and $\alpha = -0.4 \pm 0.1$.
Positional uncertainties for both sources are $\approx0\farcs3$ in
each coordinate.
No emission is detected from either source in Stokes $Q$, $U$ or $V$.

\section{Discussion}
\label{sec_discuss}

The source \snr\ is significantly linearly polarized and has a non-thermal
spectrum. Furthermore, the source has a distinct shell
morphology, and shows no significant
counterpart in 60 $\mu$m {\em IRAS}\ data.
These are all the characteristic properties of supernova remnants 
(e.g. \cite{wg96}),
and we thus classify \snr\ as a previously unidentified SNR.

\subsection{Physical Properties of \snr}
\label{sec_age}

Distances to SNRs are notoriously difficult to determine.  The
purported $\Sigma-D$ relation has extremely large uncertainties, and
this source is most likely too faint to show \HI\ absorption. So while
we cannot determine a distance to \snr\ directly, we can attempt to estimate its
distance by associating it with other objects in the region. Indeed
hydrogen recombination lines from extended thermal material have been
detected from the direction of \snr\ (\cite{lph96}), at systemic
velocities of $+42$ and $+99$~\kms. 
Adopting a standard model for Galactic rotation (\cite{fbs89}),
these velocities correspond to possible distances of 3, 6, 9 or 12~kpc,
a result which is not particularly constraining.

Nevertheless,
\snr\ is of sufficiently small angular size that we can put an upper limit
on its age simply by assuming that it is located within the Galaxy. At
a maximum distance of 20~kpc, the SNR is $27.5\pm1.5$~pc across. For a
uniform ambient medium of density $n_0$~cm$^{-3}$, the SNR has then
swept up $(260\pm40)n_0$~$M_{\sun}$ from the ISM which, for typical ejected
masses and ambient densities, corresponds to a SNR which has almost
completed the transition from free expansion to the adiabatic
(Sedov-Taylor) phase (see e.g.\ \cite{dj96}).  Thus expansion
in the adiabatic phase acts as an upper limit, and
we can derive a maximum age for
\snr\ of $(13\pm4) \left(n_0/E_{51} \right)^{1/2}$~kyr,
where $E_{51}$ is the kinetic energy of the explosion in units of
$10^{51}$~erg. For a typical value $n_0/E_{51}=0.2$ (\cite{fgw94}), we find
that the age of the SNR must be less than 8~kyr. For distances nearer than
20~kpc, the SNR is even younger. For example, at a distance of
10~kpc, the SNR has swept up sufficiently little material from the ISM
that it is still freely expanding, and an expansion velocity
5000~\kms\ then corresponds to an age 1.4~kyr.

\subsection{An association with \axp?}

\snr\ is a young SNR in the vicinity of a slow X-ray pulsar.  If the
two can be shown to be associated, and if we assume that \axp\ was
born spinning rapidly, then the youth of the system argues that \axp\
has slowed down to its current period via electromagnetic braking rather
than accretion torque, and that it is thus best interpreted as
a magnetar (cf.  \cite{vg97}). Indeed if one assumes that the source has
slowed down through the former process, its inferred dipole magnetic field
is $\sim9t_3^{-1/2} \times 10^{14}$~G, for an age
$t_3$~kyr. For ages in the range 1.4--8~kyr (\S\ref{sec_age}
above), this results in a field in the range $(3-8)\times10^{14}$~G,
typical of other sources claimed to be magnetars.

But are the two sources associated?  Associations between neutron
stars and SNRs are judged on various criteria, including agreements in
distance and in age, positional coincidence, and evidence for interaction.
Age and distance are the most fundamental of these, but unfortunately
existing data on \axp\ provide no constraints on an age, and suggest
only a very approximate distance of $\sim$10~kpc (GV98; T98).

The source \ax2\ (\cite{vgtg99}) is located well within the confines
of \snr, less than $40\arcsec$ from the center of the remnant (see
Figure~\ref{fig_snr}).  Vasisht et al. (1999) argue that \axp\ and
\ax2\ are the same source; if we assume that this source is associated
with the SNR and was born at the remnant's center, then we can infer an
upper limit on its transverse velocity of $1900d_{10}/t_3$~\kms, where
the distance to the system is $10d_{10}$~kpc.  In \S\ref{sec_age} we
estimated $d_{10}/t_3 \sim 0.3 - 0.7$, and so the velocity inferred
falls comfortably within the range seen for the radio pulsar population
(e.g.\ \cite{ll94}; \cite{cc98}) Alternatively, if we assume a
transverse velocity of $400v_{400}$~\kms, we can infer an age
for the system of $<5d_{10}/v_{400}$~kyr, consistent with the
determinations above.  There is no obvious radio counterpart to the
X-ray pulsar --- both radio point sources in the region are outside all
of the X-ray error circles.  At the position of
\ax2, we set a 5$\sigma$ upper limit of
1~mJy on the 5~GHz flux density of any point source.


We also need to consider the possibility that the positional alignment
of \ax2\ and \snr\ is simply by chance.  The region 
is a complex part of the Galactic Plane --- there are 15
catalogued SNRs within 5\arcdeg --- and it seems reasonable in such
a region that unassociated SNRs and neutron stars could lie along the
same line of sight (\cite{gj95c}).  Many young radio pulsars have no
associated SNR (\cite{bgl89}), so there is no reason to demand that 
even a young neutron star be associated with a SNR.

The first quadrant of the Galaxy is not well-surveyed for SNRs, so
we estimate the likelihood of a chance association by considering the
fourth quadrant, which has been thoroughly surveyed for SNRs by Whiteoak
\& Green (1996)\nocite{wg96}. In a representative region of the sky
defined by $320^{\circ} \le l \le 355^{\circ}$ and $|b|\le1.5^{\circ}$,
we find 44 SNRs in their catalogue.  Thus for the $\sim$10 radio-quiet
neutron stars, AXPs and SGRs at comparable longitudes and latitudes,
there is a probability $1.6\times10^{-3}$ that at least one will lie
within $40\arcsec$ of the center of a known SNR by chance.  Of course in
the present case we have carried out a targeted search towards a given
position, and so the probability of spatial coincidence is somewhat
higher than for a survey; nevertheless, we regard it unlikely
that \ax2\ should lie so close to the center of an unrelated SNR, and
hence propose that the pulsar and the SNR are genuinely associated.

There is good evidence that magnetars power radio synchrotron
nebulae through the injection of relativistic particles into their
environment (\cite{kfk+94}; \cite{fkb99}). The two such sources known
are filled-center nebulae with spectral indices $\alpha \sim -0.7$,
and in one case the neutron star is substantially offset from the core
of its associated nebula (\cite{hkc+99}).  In Figure~\ref{fig_snr},
the clump of emission with peak at $18^{\rm h}44^{\rm m}56^{\rm
s}$, $-02\arcdeg57\arcmin$ (J2000) has such properties, and one
can speculate that it corresponds to such a source.  Alternatively,
compact steep-spectrum features are seen in other shell SNRs, and may be
indicative of deceleration of the shock in regions where it is expanding
into a dense ambient medium (\cite{dbwg91}; \cite{gbm+98}).

\section{Conclusions}

Radio observations of the field of the slow X-ray pulsar \axp\ reveal
a linearly polarized non-thermal shell, \snr, which we classify as
a previously undiscovered supernova remnant. We infer that \snr\ is
young, with an upper limit on its age of 8000~yr. The proposed quiescent
counterpart of \axp, \ax2, is almost at the center of \snr, from which
we argue that the pulsar and SNR were created in the same supernova
explosion. The young age of the system provides further evidence that
anomalous X-ray pulsars are isolated magnetars rather than accreting
systems, although we caution that the apparent flux variability
of \axp\ raises questions over
both its classification as an AXP and its
positional coincidence with \snr.
Future X-ray measurements should be able to clarify the situation.


There are now six known AXPs, three of which have been
associated with SNRs. 
In every case the pulsar is at or near the
geometric center of its SNR. This result is certainly consistent with
AXPs being young, isolated neutron stars, as argued by the magnetar
hypothesis.  If one considers the radio pulsar population, the fraction
of pulsars younger than a given age which can be convincingly
associated with SNRs drops as the age threshold increases. 
The age below
which 50\% of pulsars have good SNR associations 
is $\sim$20~kyr, and for
several of these the pulsar is significantly offset from the center of 
its SNR (e.g.\ \cite{fk91}; \cite{fggd96}).  Thus if the SNRs associated
with both AXPs and radio pulsars come from similar explosions and
evolve into similar environments, this seems good evidence that AXPs
are considerably younger than 20~kyr. 
Indeed all of the three SNRs associated with AXPs have
ages $<$10~kyr (\cite{gv97};
\cite{pof+98}; \S\ref{sec_age} of this paper).  While the sample of
AXPs is no doubt incomplete, this implies a Galactic 
birth-rate for AXPs of $>$0.6 kyr$^{-1}$. This corresponds to
$(5\pm2)$\% of core-collapse supernovae (\cite{ctt+97}), or
3\%--20\% of the radio pulsar population (\cite{lml+98}; \cite{bj98}).

There is mounting evidence that soft $\gamma$-ray repeaters (SGRs) are
also magnetars (\cite{ksh+99}). However of the four known SGRs, two
(0526--66 and 1627--41) are on the edge of young SNRs (\cite{cdt+82};
\cite{sbl99}), a third (1900+14) is on the edge of an old SNR
(\cite{vkfg94}), and the fourth (1806--20) has no associated SNR blast
wave (\cite{kfk+94}).  This suggests that SGRs represent an older, or
higher velocity, manifestation of magnetars than do AXPs.

\begin{acknowledgements}

B.M.G. thanks Bob Sault for advice on calibration.
The National Radio Astronomy Observatory is a facility of the National
Science Foundation operated under cooperative agreement by Associated
Universities, Inc.  
B.M.G.  acknowledges the support of NASA through Hubble Fellowship grant
HF-01107.01-98A awarded by the Space Telescope Science Institute, which
is operated by the Association of Universities for Research in
Astronomy, Inc., for NASA under contract NAS 5--26555.
E.V.G. \& G.V.'s research is supported by NASA LTSA grant NAG5--22250

\end{acknowledgements}




\vspace{-20mm}

\begin{deluxetable}{lccc}
\tablewidth{400pt}
\tablecaption{VLA observations of \snr.\label{tab_obs}}
\tablehead{
\colhead {} & \colhead{5~GHz} & \colhead{8~GHz} & \colhead{}  }
\startdata
Resolution  & $12\farcs8 \times 12\farcs5$ & $8\farcs2 \times 7\farcs5$ \nl
rms noise ($\mu$Jy~beam$^{-1}$) & 100 & 80 \nl
Flux density of \snr\ (mJy)\tablenotemark{a}  & $410\pm5$ & $260\pm10$ \nl
Center of \snr\ & \multicolumn{3}{l}{$18^{\rm h}44^{\rm m}52^{\rm s}$, 
$-02\arcdeg56\arcmin30\arcsec$ ($\alpha$, $\delta$; J2000)} \nl
   & \multicolumn{3}{l}{29.57, +0.12 ($l$, $b$)} \nl
Diameter of \snr\  & \multicolumn{2}{l}{$4\farcm5 \times 5\farcm0$} \nl
\enddata
\vspace{-2mm}
\tablenotetext{a}{These flux densities are underestimates, since the
largest spatial scales were not well-imaged.}
\end{deluxetable}

\end{document}